\documentclass[a4paper,twoside]{article}
\usepackage{epsfig}
\usepackage{subfigure}

\usepackage{graphicx}
\graphicspath{{\getgraphicspath}}
\newcommand{\getgraphicspath}{graphics/}
\usepackage[hyperref,dvipsnames]{xcolor}
\usepackage{tikz}
\usepackage{epstopdf}
\usepackage{pgfplots}
\usepackage{fancybox}
\usepackage{lipsum}
\usepackage[many]{tcolorbox}    
\usepackage{wrapfig}
\usepackage{calc}
\usepackage{amssymb}
\usepackage{amstext}
\usepackage{amsmath}
\usepackage{amsthm}
\usepackage{multicol}
\usepackage{pslatex}
\usepackage{natbib}
\let\bibhang\relax
\usepackage{apalike}
\usepackage{rotating}
\usepackage{url}
\usepackage{SCITEPRESS}     

\begin{document}

\title{Motivations, Classification and Model Trial of Conversational Agents for Insurance Companies}

\author{\authorname{Falko Koetter\sup{1}, Matthias Blohm\sup{1}, Monika Kochanowski\sup{1}, Joscha Goetzer\sup{2}, Daniel Graziotin\sup{2},\newline and Stefan Wagner\sup{2}}
\affiliation{\sup{1}Fraunhofer Institute for Industrial Engineering, Nobelstr. 12, 70569 Stuttgart, Germany}
\affiliation{\sup{2}University of Stuttgart, Universit\"atsstr. 38, 70569 Stuttgart, Germany\\}
\email{falko.koetter@iao.fraunhofer.de, matthias.blohm@iao.fraunhofer.de, monika.kochanowski@iao.fraunhofer.de, joscha.goetzer@gmail.com, daniel.graziotin@iste.uni-stuttgart.de, stefan.wagner@iste.uni-stuttgart.de}
}


\abstract{Advances in artificial intelligence have renewed interest in conversational agents. So-called chatbots have reached maturity for industrial applications. German insurance companies are interested in improving their customer service and digitizing their business processes. In this work we investigate the potential use of conversational agents in insurance companies by determining which classes of agents are of interest to insurance companies, finding relevant use cases and requirements, and developing a prototype for an exemplary insurance scenario. Based on this approach, we derive key findings for conversational agent implementation in insurance companies.}

\onecolumn \maketitle \normalsize \vfill

\section{\uppercase{Introduction}}
\label{sec:introduction}

\noindent With the digital transformation changing usage patterns and consumer expectations, many industries need to adapt to new realities. The insurance sector is next in line to grapple with the risks and opportunities of emerging technologies, in particular \emph{Artificial Intelligence}~\citep{aichanging}. 

Fraunhofer~IAO as an applied research institution supports digital transformation processes in an ongoing project with multiple insurance companies~\citep{iao2018innonetz}. The goal of this project is to scout new technologies, investigate them, rate their relevance and evaluate them (e.g. in a model trial or by implementing a prototype). While insurance has traditionally been an industry with very low customer engagement, insurers now face a young generation of consumers with changing attitudes regarding insurance products and services~\citep{iao2017zukunftsstudie}. 

Traditionally, customer engagement uses channels like mail, telephone and local agents. In 2016, chatbots emerged as a new trend~\citep{customerservice}, making it a topic of interest for Fraunhofer~IAO and insurance companies.

With the rise of the smartphone, many insurers started offering apps, but success was limited~\citep{jdpower2017}, which may stem from app fatigue~\citep{appfatigue}. App use has plateaued, as users have too many apps and are reluctant to add more~\citep{gartner}. In contrast, conversational~agents require no separate installation, as they are accessible via messaging apps, which are likely to be already installed on a user's smartphone. Conversational agents are an alternative to improve customer support and digitize processes like claim handling.

The objective of this work is to facilitate the creation of conversational agents by defining the traits of an agent more clearly using a (1) classification framework, which is based on current literature and research topics, and systematically analyzing (2) use cases and requirements in an industry, shown in the example insurance scenario. The applicability of this approach is shown by implementing a prototype (3) including an evaluation. Furthermore, we derive key findings for conversational agent implementation in insurance companies and open points for research.

\section{\uppercase{Related Work}}
\label{sec:relatedwork}

\noindent In this section we investigate work in the area of conversational agents, dialog management, and research applications in insurance.

\cite{theconversationalinterface} offer detailed explanations about background and history of conversational interfaces as well as techniques to build and evaluate own agent applications. Another literature review about chatbots was provided by \cite{chatbotthesis}, where common approaches and design choices are summarized followed by a case study about the functioning of IBM's chatbot Watson, which became famous for winning the popular quiz game \emph{Jeopardy!} against humans.

Many chatbot applications have already been built nowadays with the goal to solve actual problems. One example is PriBot, a conversational agent, which can be asked questions about an application's privacy policy, because users tended to skip reading the often long and difficult to understand privacy notices. Also, the chatbot accepts queries of the user which aim to change his privacy settings or app permissions~\citep{pribots}.

In the past there have already been several studies with the goal to evaluate how a conversational agent should behave for being considered as human-like as possible. In one of them, conducted by~\cite{hallmarks}, fourteen participants were asked to talk to an existing chatbot and to collect key points of convincing and unconvincing characteristics. It turned out that the bot's ability to hold a theme over a longer dialog made it more realistic. On the other hand, not being able to answer to a user's questions was regarded as an unsatisfying characteristic of the artificial conversational partner~\citep{hallmarks}.

In another experiment, which was done by ~\cite{onboarding}, eight users had to talk to two different kinds of chatbots, one behaving more human-like and one behaving more robotic. In this context, they had to fulfill certain tasks like ordering an insurance policy or demanding an insurance certification. All of the participants instinctively started to chat by using natural human language. In cases in which the bot did not respond to their queries in a satisfying way, the users' sentences continuously got shorter until they ended up with writing key words only. Thus, according to the results of this survey, conversational agents preferably should be created human-like, because users seem to be more comfortable when feeling like talking to another human being, especially in cases in which the concerns are crucial topics like their insurance policies~\citep{onboarding}. 

\emph{Dialog management strategies} (DM) define the conversational behaviors of a system in response to user message and system state~\cite{theconversationalinterface}. 

In industry applications, DM often consists of a handcrafted set of rules and heuristics, which are tightly coupled to the application domain \citep{theconversationalinterface} and improved iteratively. One problem with handcrafted approaches to DM is that it is challenging to anticipate every possible user input and react appropriately, making development resource-intensive and error-prone. But if few or no recordings of conversations are available, these \emph{rule-oriented} strategies may be the only option.

As opposed to the rule-oriented strategies, data-oriented architectures work by using machine learning algorithms that are trained with samples of dialogs in order to reproduce the interactions that are observed in the training data. 
These statistical or heuristical approaches to DM can be classified into three main categories: Dialog modeling based on \emph{reinforcement learning}, \emph{corpus-based} statistical dialog management, and \emph{example-based} dialog management (simply extracting rules from data instead of manually coding them)~\citep{theconversationalinterface, spierling2005interactive}. \cite{spierling2005interactive} highlights neural networks, Hidden-Markov Models, and  Partially Observable Markov Decision Processes as possible implementation technologies.

The following are common strategies for rule-based dialog management:

\begin{itemize}

\item Finite-state-based DM uses a finite state machine with handcrafted rules, and performs well for highly structured, system-directed tasks~\citep{theconversationalinterface}.
\item Frame-based DM follows no predefined dialog path, but instead allows to gather pieces of information in a frame structure and no specific order. This is done by adding an additional entity-value slot for every piece of information to be collected and by annotating the intents in which they might occur. Using frames, a less restricted, user-directed conversation flow is possible, as data is captured as it comes to the mind of the user~\citep{rudnicky1999agenda}.
\item Information State Update represents the information known at a given state in a dialog and updates the internal model each time a participant performs a \emph{dialog move}, (e.g. asking or answering). The state includes information about the mental states
of the participants (beliefs, desires, intentions, etc.) and about the dialog (utterances, shared information, etc.) in abstract representations. Using so-called update moves, applicable moves are chosen based on the state~\citep{traum2003isu}.
\item Agent-based DM uses an  agent that fulfills conversation goals by dynamically using plans for tasks like intent detection and answer generation. The agent has a set of beliefs and goals as well as an information base which is updated throughout the conversation. Within this information framework the agent continuously prioritizes goals and autonomously selects plans that maximize the likelihood of goal fulfillment~\citep{nguyen2005agent}.

\end{itemize}

\cite{agentdm} describes how multiple DM approaches can be combined to use the best strategy for specific circumstances.

A virtual insurance conversational agent is described by \cite{yacoubi2018teatime}, utilizing \emph{TEATIME}, an architecture for agent-based DM. TEATIME uses emotional state as a driver for actions, e.g. when the bot is perceived unhelpful, that emotion leads the bot to apologize. The shown example bot is a proof of concept for TEATIME capable of answering questions regarding insurance and react to customer emotions, but does not implement a full business process.

\cite{kowatsch2017text} describe a text-based healthcare chatbot that acts as a companion for weightloss but also connects a patient with healthcare professionals. The chat interface supports non-textual inputs like scales and pictorials to gather patient feedback. Study results showed a high engagement with the chatbot as a peer and a higher percentage of automated conversation the longer the chatbot is used.

Overall, these examples show potential for conversational agents in the insurance area, but lack support for complete business processes.

\section{\uppercase{Classification of Conversational Agents}}
\label{sec:agents}

\noindent The idea of conversational agents that are able to communicate with human beings is not new: In 1966, Joseph Weizenbaum introduced \emph{Eliza}, a virtual psychotherapist, which was able to respond to user queries using natural language and which could be considered as the first \emph{chatbot}~\citep{eliza1966}. Nowadays, the idea of speaking machines has experienced a revival with the emergence of new technologies, especially in the area of artificial intelligence. Novel machine learning algorithms allow developers to create software agents in a much more sophisticated way and in many cases they already outperform previous statistical NLP methods~\citep{theconversationalinterface}. Additionally, the importance of messaging apps such as WhatsApp or Telegram has increased over the last years. In 2015, the total number of people using these messaging services outran the total number of active users in social networks for the first time. Today, each of these app has about between 200 million and 1.5 billion users~\citep{messengers}. 

As a highly popular topic in 2016~\citep{customerservice}, a great variety of different chatbots evolved together with an equally wide range of terminologies. For being able to draw a big picture of the current trends in the area of conversational agents, we divide them into the following four common categories:

\begin{itemize}
	\item \textbf{Chatterbots}: Bots with focus on small talk and realistic conversations, not task-oriented, e.g. Cleverbot~\citep{cleverbot}.
	\item \textbf{(Virtual, Intelligent, Cognitive, Digital, Personal) assistants
  (VPAs)}: Agents fulfilling tasks intelligently based on spoken or written user input and with the help of data bases and personalized user preferences~\citep{cooper2008personal} (e.g. Apple's Siri or Amazon's Alexa~\citep{chatbotsreturn}).
	\item \textbf{Specialized digital assistants (SDAs)}: Focused on a specific domain of expertise, goal-oriented behavior ~\citep{chatbotsreturn}.
	\item \textbf{Embodied conversational agents (ECAs)}: Visually animated agents, e.g. in form of avatars or robots~\citep{evaluating2017}, where speech is combined with gestures and facial expressions.
\end{itemize}

Figure \ref{fig:cui-types} shows the results of evaluating these four classes in terms of different characteristics such as \emph{realism} or \emph{task orientation}based on own literature research. Chatterbots provide a high degree of entertainment since they try to imitate the behavior of human beings while chatting, but there is no specific goal to be reached within the scope of these conversations. In contrast, general assistants like Siri or Alexa are usually called by voice in order to fulfill a specific task. Specialized assistants concentrate even more on achieving a specific goal, which often comes at the expense of realism and user amusement because their ability to respond to not goal-oriented conversational inputs like small talk is mostly limited. The best feeling of companionship can be experienced by talking to an embodied agent, since the reactions of these bots are closest to human-like behavior.

When looking at these classification results, a broad spectrum of various possible agents is offered. Therefore, a restriction depending on the specific use case has to be made first, before the realization of a prototypical chatbot can be tackled. Since Fraunhofer IAO aims to investigate solutions supporting processes in the insurance domain, creating a prototype with the properties of a SDA is necessary, because the main purpose in this scenario is to perform and successfully complete a certain task (e.g. reporting a claim). Furthermore, adding additional chatterbot features such as the ability to do small talk in a limited goal-oriented scope could lead to a more realistic and human-like user experience. However, before considering detailed design and implementation choices, it is helpful to take a general look at the role of chatbots within the special environment of insurance companies for identifying essential issues and needs.

\usetikzlibrary{shapes}

\newcommand{\DU}{6} 
\newcommand{\UU}{7} 

\newdimen\R 
\R=3.5cm 
\newdimen\L 
\L=4.1cm

\newcommand{\AW}{360/\DU} 



	
	\begin{flushleft}

    \begin{tikzpicture}[scale=0.5]
      \path (0:0cm) coordinate (O); 
			
      \foreach \X in {1,...,\DU}{
        \draw (\X*\AW:0) -- (\X*\AW:\R);
        \draw (\X*\AW:0) -- (\X*\AW:\R);
      }

      \foreach \Y in {0,...,\UU}{
        \foreach \X in {1,...,\DU}{
          \path (\X*\AW:\Y*\R/\UU) coordinate (D\X-\Y);
          \fill (D\X-\Y) circle (1pt);
        }
        \draw [opacity=0.3] (0:\Y*\R/\UU) \foreach \X in {1,...,\DU}{
            -- (\X*\AW:\Y*\R/\UU)
        } -- cycle;
      }

      \path (1*\AW:\L) node (L1) {\footnotesize \textit{\textsf{Companionship}}};
      \path (2*\AW:\L) node (L2) {\footnotesize \textit{\textsf{Realism}}};
      \path (3*\AW:\L) node (L3) {\footnotesize \textit{\textsf{Entertainment}}~~~~~~~~~~~~~~~~~};
      \path (4*\AW:\L) node (L4) {\footnotesize \textit{\textsf{Textual}}};
      \path (5*\AW:\L) node (L5) {\footnotesize \textit{\textsf{Task Orientation}}};
      \path (6*\AW:\L) node (L6) {\footnotesize ~~~~~\textit{\textsf{Spoken}}};

      \newcommand{\bgc}{null}
      %
      %
      \newcommand{\opazetee}{0.35}
      \newcommand{\linewidthall}{2.1pt}

      \renewcommand{\bgc}{Magenta}
      \draw [color=\bgc,line width=\linewidthall,opacity=\opazetee,fill=\bgc]
        (D1-4) --
        (D2-6) --
        (D3-7) --
        (D4-7) --
        (D5-1) --
        (D6-2) -- cycle;
        \node[text width=5cm,color=\bgc] at (-1.8,-0.8) 
            {\scriptsize \textbf{Chatterbots}};

      \renewcommand{\bgc}{Emerald}
      \draw [color=\bgc,line width=\linewidthall,opacity=\opazetee,fill=\bgc]
        (D1-3) -- 
        (D2-2) --
        (D3-3) --
        (D4-2) --
        (D5-6) --
        (D6-7) -- cycle;
        \node[text width=5cm,color=\bgc] at (8.55,-1.0) 
            {\scriptsize \textbf{General\\Digital Assistants}};

      \renewcommand{\bgc}{RoyalPurple}
      \draw [color=\bgc,line width=\linewidthall,opacity=\opazetee,fill=\bgc]
        (D1-7) --
        (D2-7) --
        (D3-4) --
        (D4-1) --
        (D5-1) --
        (D6-5) -- cycle;
        \node[text width=5cm,color=\bgc] at (7.55,2.3) 
            {\scriptsize \textbf{Embodied\\Conversational Agents}};

      \renewcommand{\bgc}{Dandelion}
      \draw [color=\bgc,line width=\linewidthall,opacity=\opazetee,fill=\bgc]
        (D1-1) --
        (D2-1) --
        (D3-2) --
        (D4-6) --
        (D5-7) --
        (D6-4) -- cycle;
        \node[text width=5cm,color=\bgc] at (7.80,-2.55) 
            {\scriptsize \textbf{Specialized\\Digital Assistants}};

       \renewcommand{\bgc}{Black}
			\node[color=Goldenrod,outer sep=1] at (D1-0) {\textbf{\small 0}};
			\node[color=Goldenrod,outer sep=1] at (D1-1) {\textbf{\small 1}};
			\node[color=Goldenrod,outer sep=1] at (D1-2) {\textbf{\small 2}};
			\node[color=Goldenrod,outer sep=1] at (D1-3) {\textbf{\small 3}};
			\node[color=Goldenrod,outer sep=1] at (D1-4) {\textbf{\small 4}};
			\node[color=Goldenrod,outer sep=1] at (D1-5) {\textbf{\small 5}};
			\node[color=Goldenrod,outer sep=1] at (D1-6) {\textbf{\small 6}};
			\node[color=Goldenrod,outer sep=1] at (D1-7) {\textbf{\small 7}};


      \end{tikzpicture}
				\end{flushleft}
   
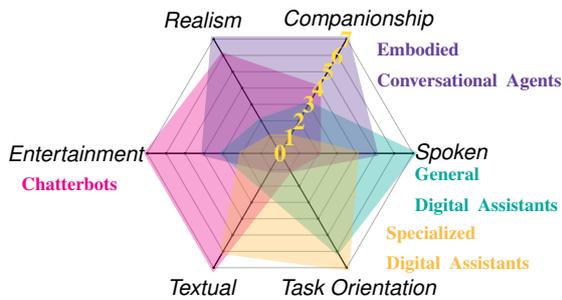
\captionof{figure}{Classification of conversational agents with their characteristics (own presentation). Values between 0 and 7 indicate how strong a characteristic applies for the given type of agent.}\label{fig:cui-types}

\section{\uppercase{Chatbots in Insurance}}
\label{sec:methodology}

\noindent Insurance is an important industry sector in Germany, with 560 companies that manage about 460 million policies~\citep{gdv2014statistik}. However, the insurance sector is under a high cost pressure, which shows in a declining employee count and low margins~\citep{stange2015zukunft}. The insurance market is saturated and has transitioned from a growth market to a displacement market~\citep{aschenbrenner2010versicherung}. For the greater part, German insurance companies have used conservative strategies, caused by risk aversion, long-lived products, hierarchical structures, and profitable capital markets~\citep{veraenderungsaversion}. As these conditions change, so must insurance companies.

Insurance is an industry with low customer engagement, as an insurer traditionally has basically two touch points to interact with customers: \emph{selling a product} and \emph{the claims process}. A study found that consumers interact less with insurers than with any other industry, so the consumer experience with insurers tends to lag behind others~\citep{engagement}. 

Many insurance companies have heterogeneous IT infrastructures incorporating legacy systems (sometimes from two or more companies as the result of a merger)~\citep{digitalenterprise}. These grown architectures pose challenges when implementing new data-driven or AI solutions, due to issues like data quality, availability and privacy. Nonetheless, the high amount of available data and complex processes make insurance a prime candidate for machine learning and data mining. The adoption of AI in the insurance sector is in early stages, but accelerating, as insurance companies strive to improve service and remain competitive~\citep{aichanging}.

Conversational agents are one AI technology at the verge of adoption. In 2017, ARAG launched a travel insurance chatbot, quickly followed by bots from other insurance companies~\citep{aragchatbot}. While these chatbots are still experimental and implement narrow use cases, these first implementations prove public interest and feasibility.

\begin{figure}[htbp]
	\includegraphics[width=0.48\textwidth]{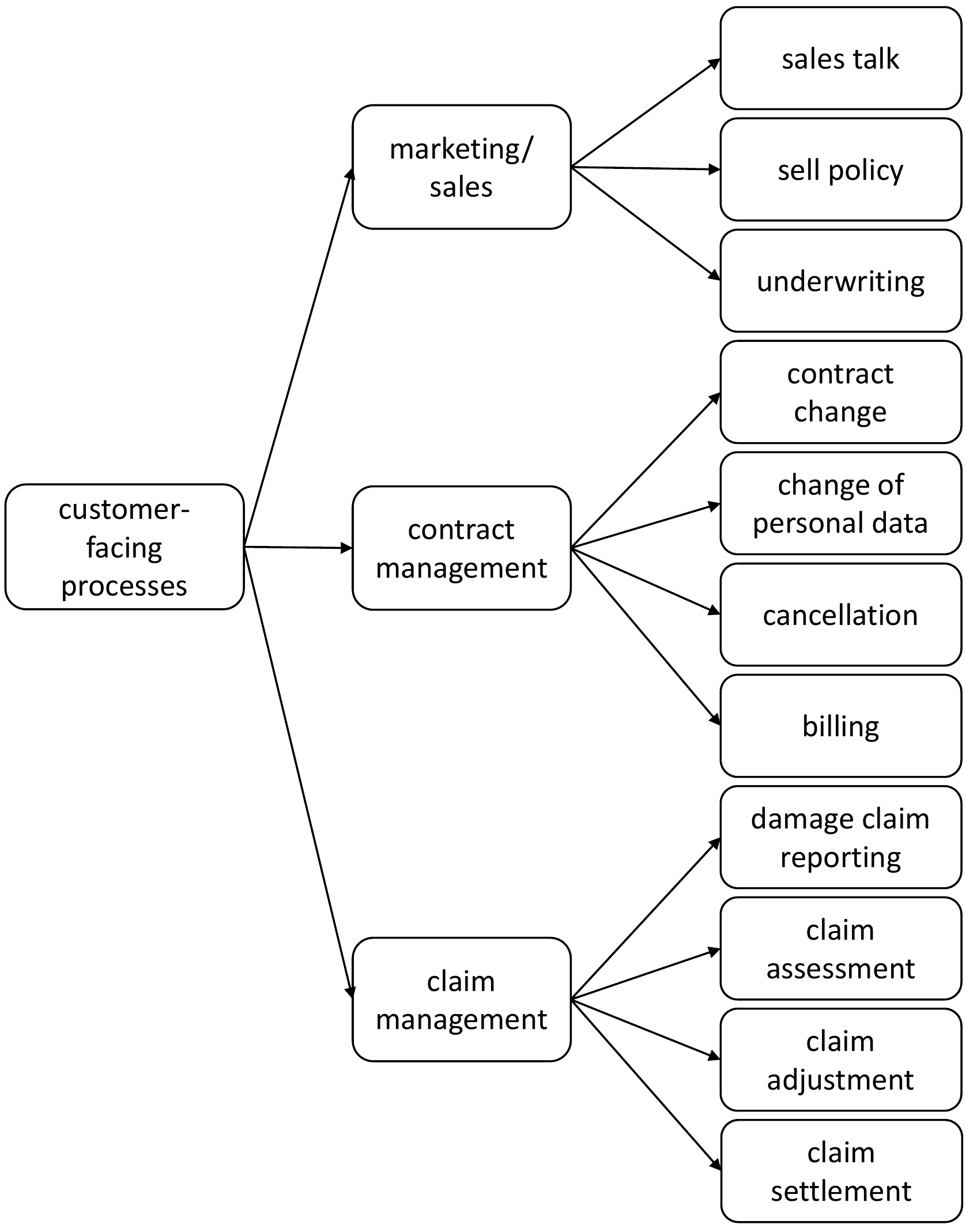}
	\centering
	\caption{Customer-facing insurance processes (based on~\cite{aschenbrenner2010versicherung} and \cite{horch2012openxchange})}
	\label{fig:processes}
\end{figure}

To identify areas of possible chatbot support, we surveyed the core business processes of insurance companies as described in ~\cite{aschenbrenner2010versicherung} and \cite{horch2012openxchange}. Three core areas of insurance companies are customer-facing: \emph{marketing/sales}, \emph{contract management} and \emph{claim management}. Figure~\ref{fig:processes} shows the main identified processes related to this area.

We identified all these processes as possible use cases for conversational agent support, in particular support by SDAs. 

Furthermore, we investigated general requirements for conversational agents in these processes:

\textbf{Availability and ease-of-use} Conversational agents are an alternative to both conventional customer support (e.g. phone, mail) as well as conventional applications (e.g. apps and websites). Compared to these conventional solutions, chatbots offer more availability than human agents and have less barriers of use than conventional applications, requiring neither an installation nor the ability to learn a new user interface, as conventional messaging services are used~\citep{derler2017}.

\textbf{Guided information flow} Compared to websites, which offer users a large amount of information they must filter and prioritize themselves, conversational agents offer information gradually and only after the intent of the user is known. Thus, the search space is narrowed at the beginning of the conversation without the user needing to be aware of all existing options.

\textbf{Smartphone integration} Using messaging services, conversational agents can integrate with other smartphone capabilities, e.g. making a picture, sending a calendar event, setting a reminder or calling a phone number.

\textbf{Customer call reduction} Customer service functions can be measured by reduction of customer calls and average handling time~\citep{customerservice}. SDAs can help here by automating conversations, handling standard customer requests and performing parts of conversations (e.g. authentication). 

\textbf{Human handover} Customers often use social media channels to escalate an issue in the expectation of a \emph{human} response, instead of an \emph{automated} one. A conversational agent thus must be able to differentiate between standard use cases it can handle and more complicated issues, which need to be handed over to human agents~\citep{risk}. One possible approach is to use sentiment detection, so customer who are already stressed are not further aggravated by a bot~\citep{customerservice}.

\textbf{Digitize claim handling} Damage claim handling in insurance companies is a complex process involving multiple departments and stakeholders~\citep{koetter2012business}. Claim handling processes are more and more digitized within the insurance companies~\citep{horch2012openxchange}, but paper still dominates communication with claimants, workshops and experts. \citep{touchless} defines maturity levels of insurance processes, defining \emph{virtual handling} as a process where claims are assessed fully digitally based on digital data from the claimant (e.g. a video, a filled digital form), and \emph{touchless handling} as a fully digital process with no human intervention on the insurance side. SDAs help moving towards these maturity levels by providing a guided way to make a claim digitally and communicate with the claimant (e.g. in case additional data is needed).

\textbf{Conversational commerce} is the use of Conversational Agents for marketing and sales related purposes~\citep{commerce}. Conversational Agents can perform multiple tasks using a single interface. Examples are using opportunities to sell additional products (\emph{cross-sell}) or better versions of the product the customer already has (\emph{up-sell}) by chiming in with personalized product recommendations in the most appropriate situations. One example would be to note that a person's last name has changed during an address update customer service case and offer appropriate products if the customer has just married.

\textbf{Internationalization} is an important topic for large international insurance companies. However, most frameworks for implementing conversational agents are available in more than one language. To the best of our knowledge today the applied conversational agents in German insurance are optimized only for one language. So this topic is future work in respect to the prototype. 

\textbf{Compliance} to privacy (GDPR) is usually guaranteed by the login mechanisms on the insurance sites, therefore the topic is out of scope for our research prototype. For broader scenarios not requiring identification on the insurance site and the usage of the data for non-costumers, this is an open area of research. 

\section{\uppercase{Prototype}}
\label{sec:prototype}

\noindent Based on the work presented in the last sections and our talks with insurance companies, we arrived at the following non-functional requirements that the chatbot prototype ideally should fulfill: 

\begin{itemize}
	\item \textbf{Interoperability}: The agent should be able to keep track of the conversational context over several message steps and messengers.
	\item \textbf{Portability}: The agent can be run on different devices and platforms (e.g. Facebook Messenger, Telegram). Therefore it should use a unified, platform-independent messaging format.
	\item \textbf{Extensibility}: The agent should provide a high level of abstraction that allows designers to add new conversational content without having to deal with complicated data structures or code.
\end{itemize}

Additionally, the following functional requirements should be regarded in the implementation:

\begin{itemize}
	\item \textbf{Report a claim}: The system must provide the possibility for a user to report a damage claim using the conversational agent (prototype scenario).
	\item \textbf{Human language understanding}: The system should be able to understand and process the user's inputs and intents given in form of written natural language (German).
	\item \textbf{Response generation}: The system should be able to generate an answer sentence in written human language (German) according to the user queries.
\end{itemize}
 
For dialog design within the prototype, experimenting with machine learning algorithms was the preferred implementation strategy. For this purpose, discussions with insurance companies were held to assess the feasibility of receiving existing dialogs with customers, for example for online chats, phone logs or similar. However, such logs generally seem to be not available at German insurers, as the industry has self-regulated to only store data needed for claim processing~\citep{wwwGDVCOC2012}. As a research institute represents a third party not directly involved in claims processing, data protection laws forbid sharing of data this way without steps to secure personal data. During our talks we have identified a need for automated or assisted anonymization of written texts as a precondition for most customer-facing machine learning use cases, at least when operating in Europe~\citep{kamarinou2016machine}. However, these issues go beyond the scope of our current project, but provide many opportunities for future research. 

To still build a demonstrator in face of these challenges, dialogs for the prototype were manually designed without using real-life customer conversations and fine-tuned by user testing with fictional damage claims. As this approach entails higher manual effort for dialog design, a narrower scenario was chosen to still allow for the full realization of a customer-facing process. The chosen scenario was a special case of the damage claim process: \emph{The user has a damaged smartphone or tablet and wants to make an insurance claim}.

\begin{figure*}[htb]
	\includegraphics[width=1.0\textwidth]{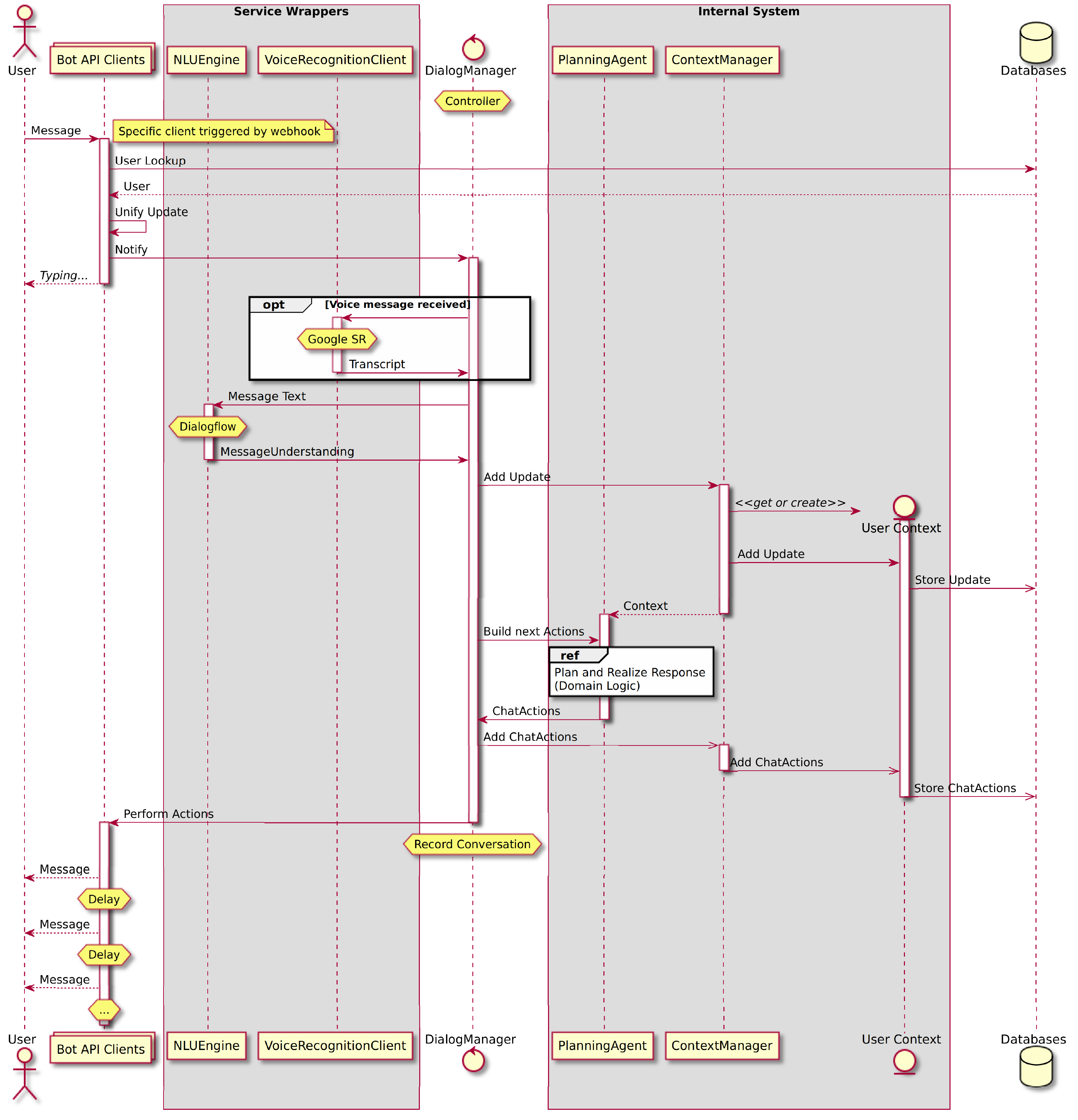}
	\centering
	\caption{Sequence diagram of the conversational agent prototype}
	\label{fig:sequencecore}
\end{figure*}

Figure~\ref{fig:sequencecore} shows the main components of the prototype and their operating sequence when processing a user message. To provide extensibility prototype architecture strictly separates service integration, internal logic and domain logic.

The user can interact with the bot over different communication channels which are integrated with different \emph{bot API clients}. To integrate a different messaging service, a new bot API client needs to be written. The remainder of the prototype can be reused.

Once a user has written a message, a lookup of \emph{user context} is performed to determine if a conversation with that user is already in progress. User context is stored in a database so no state is kept within external messaging services. Afterwards, a \emph{typing} notification is given to the user, indicating the bot has received the message and is working on it. This prevents multiple messages by a user who thinks the bot is not responsive.

In the next step, the message has to be understood by the bot. In case of a voice message, it is transcribed to text using a \emph{Google} speech recognition web service. 

For natural language understanding, we compared four possible frameworks (Microsoft's LUIS, Google's Dialogflow, Facebook's wit.ai and IBM's Watson) regarding important criteria for prototype implementation. A comparison table for these criteria is shown in Table~\ref{tab:comparison}. As a result of the comparison, Dialogflow was chosen as a basic framework.

\begin{table}[h]
  \centering
  \caption{Comparison of Microsoft's LUIS, Google's Dialogflow, Facebook's
    wit.ai, and IBM's Watson (based on~\cite{nlucomparison})} \centering
  \label{tab:comparison}
	\small
  \begin{tabular}{lllll}
    & \begin{sideways}\textbf{LUIS}\end{sideways} &  \begin{sideways}\textbf{Dialogflow}\end{sideways} & \begin{sideways}\textbf{Wit.ai}\end{sideways} & \begin{sideways}\textbf{Watson}\end{sideways} \\
    Python bindings                          & no             & yes & yes           & yes \\
    German language                          & yes            & yes                                              & in Beta         & yes        \\
    Free service                             & no             & yes                                              & yes             & no         \\
    Remember state & yes            & yes                                              & yes             & yes        \\
    Service bound                            & yes            & yes                                              & yes             & yes        \\
    Simple training                          & with effort         & yes                                              & yes             & yes       
  \end{tabular}
\end{table}

Dialogflow is used for intent identification, which determines the function of a message and based on that a set of possible parameters~\citep{theconversationalinterface}. For example, the intent of the message ``the display of my smartphone broke'' may have the intent \texttt{phone\_broken} with the parameter \texttt{damage\_type} as \texttt{display\_damage}, while the parameter \texttt{phone\_type} is not given. Together, this information given by Dialogflow is a \texttt{MessageUnderstanding}

\begin{figure*}[htb]
	\includegraphics[width=0.8\textwidth]{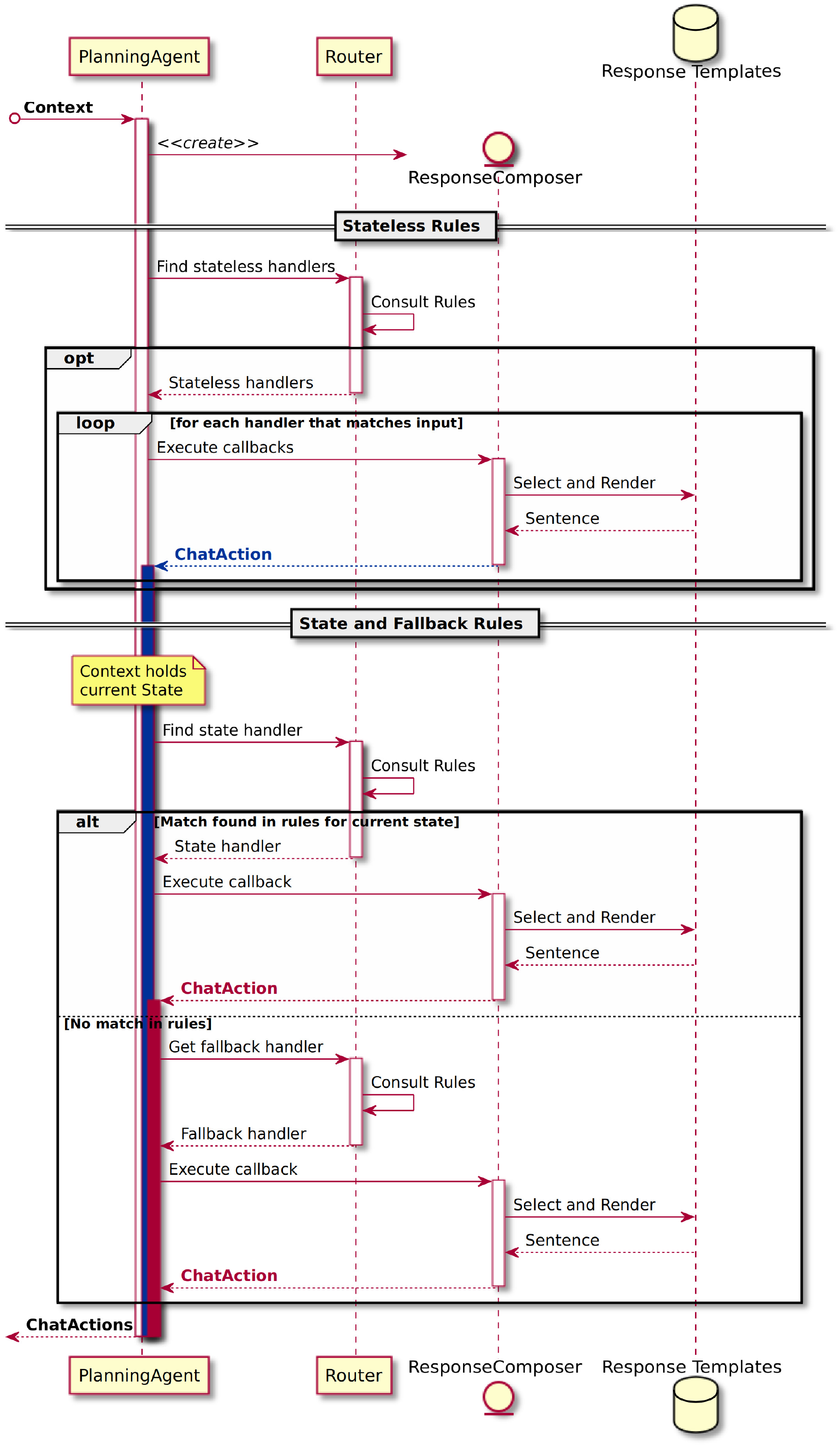}
	\centering
	\caption{Detailed sequence diagram of the response generation in the conversational agent prototype}
	\label{fig:sequencedomain}
\end{figure*}

As soon as the message is understood, the user context is updated. Afterwards, a response needs to be generated. This process, which was labeled with \emph{Plan and Realize Response} in Figure~\ref{fig:sequencecore}, is shown in detail in Figure~\ref{fig:sequencedomain}.

\begin{figure}[htbp]
	\includegraphics[width=0.45\textwidth]{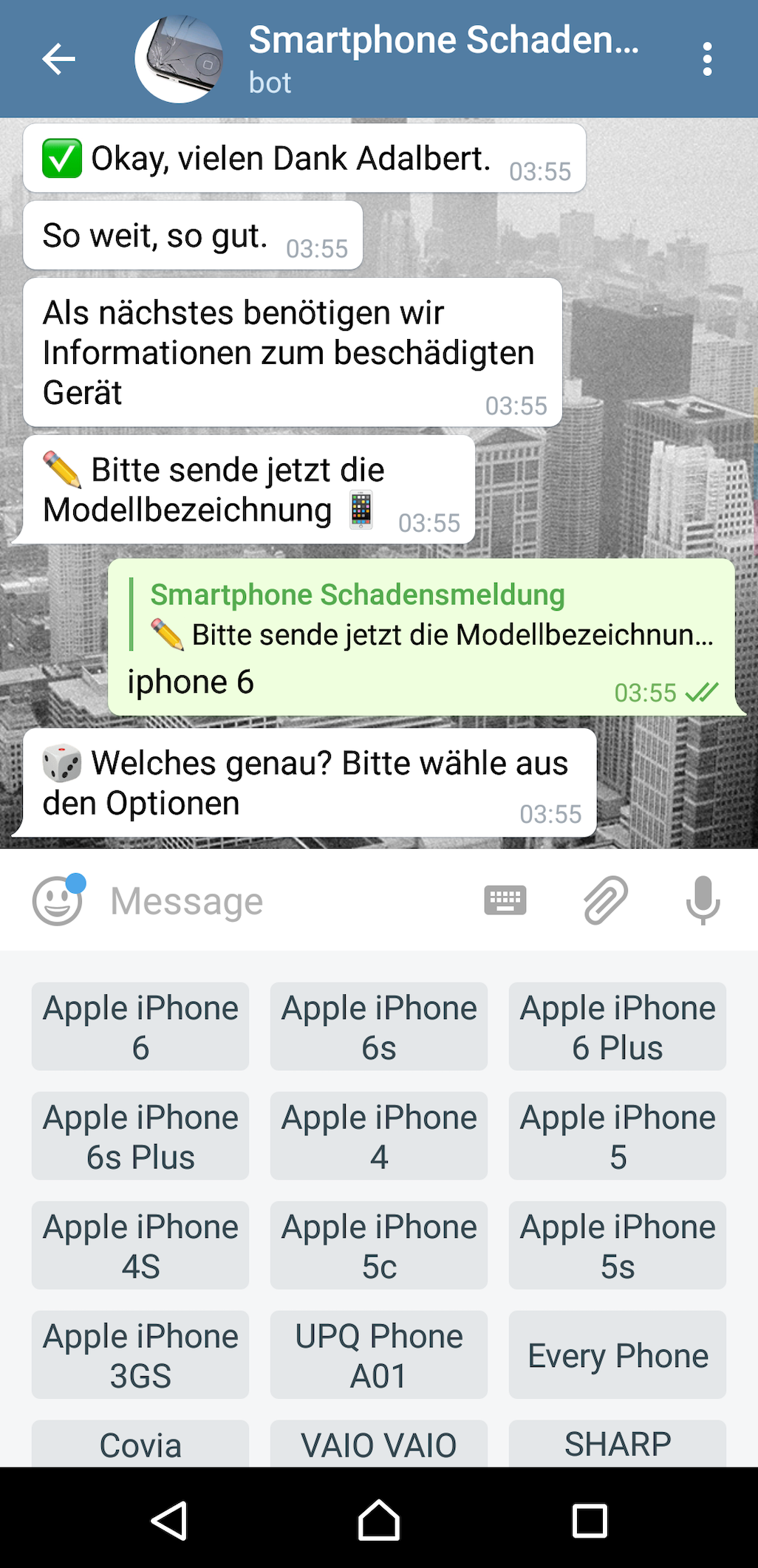}
	\centering
	\caption{Dialog excerpt of the prototype, showing the possibility to clarify the phone model via multiple-choice input)}
	\label{fig:prototypescreenshot}
\end{figure}

In the prototype, an agent-based strategy was chosen in order to combine the capabilities of the frame-based entities and parameters in Dialogflow with a custom dialog controller based on predefined rules in a finite state machine. This machine allows to define rules that trigger \emph{handlers} and \emph{state transitions} when a specific intent or entity-parameter combination is encountered. That way, both intent and frame processing happen in the same logically encapsulated unit, enabling better maintainability and extensibility. The rules are instances of a set of \texttt{*Handler} classes such as an \texttt{IntentHandler} for the aforementioned intent and parameter matching, supplemented by other handlers, e.g. an \texttt{AffirmationHandler}, which consolidates different intents that all express a confirmation along the lines of ``yes'', ``okay'', ``good'' and ``correct'', as well as a \texttt{NegationHandler}, a \texttt{MediaHandler} and an \texttt{EmojiSentimentHandler} (to analyze positive, neutral, or negative sentiment of a message with emojis). Each implements their own~ \texttt{matches(MessageUnderstanding)} method.

Rules (handlers) are used within the dialog state machine:

\begin{enumerate}
\item \emph{Stateless} handlers are checked independently of the current state. For example, a \texttt{RegexHandler} rule determines whether the formality of the address towards the user should be changed (German differentiates the informal ``du'' and the formal ``Sie'')
\item \emph{Dialog States} map each possible state to a list of handlers that are applicable in that state. For instance, when the user has given an answer and the system asks for \emph{explicit confirmation} in a state \texttt{USER\_CONFIRMING\_ANSWER}, then an \texttt{AffirmationHandler} and a \texttt{NegationHandler} capture ``yes'' and ``no'' answers.
\item \label{itm:fallbacks} \emph{Fallback} handlers are checked if none of the applicable state handlers have yielded a match for an incoming \texttt{MessageUnderstanding}. These \texttt{fallbacks} include static, predefined responses with lowest priority (e.g. small talk), as well as handlers to repair the conversation by bringing the user back on track or changing the topic.
\end{enumerate}

At first, the system had only allowed a single state to be declared at the same time in the router. However, this had quickly proven to be insufficient as users are likely to want to respond or refer not only to the most recent message, but also to previous ones in the chat. With only a single contemporaneous state, the user's next utterance is always interpreted only in that state. In order to make this model resilient, every state would need to incorporate every utterance that the user is likely to say in that context. As this is not feasible, the prototype has state handlers that allow layering transitions on top of each other, allowing multiple simultaneous states which may advance individually.

To avoid an explosion of active states, the system has \emph{state lifetimes}: new states returned by callbacks may have a lifetime that determines the number of dialog moves this state is valid for. On receiving a new message, the planning agent decreases the lifetimes of all current dialog states by one, except for the case of utter non-understanding (``fallback'' intent). If a state has exceeded its lifetime, it is removed from the priority queue of current dialog states. 

Figure~\ref{fig:sequencedomain} contains details about how the system creates responses to user queries. Based on the applicable rule, the conversational agent performs chat actions (e.g. sending a message), which are generated from response templates, taking into account dialog state, intent parameters, and information like a user's name, mood and preferred level of formality. 

RuleHandlers, states and other dialog specific implementations are encapsulated, so a new type of dialog can be implemented without needing to change the other parts of the system.

Generated chat actions are stored in the user context and performed for the user's specific messenger using the \emph{bot API}. As the user context has been updated, the next message by the user continues the conversation.

Before the user utters an intent to make a damage claim, the prototype explains its functionality and offers limited small talk. As soon as the user wants to make a damage claim, the conversational agent gathers the required information using a predetermined questionnaire. Questions concern type of damage, damaged phone, phone number, IMEI, damage time, damage event details, etc. Answers are interpreted using dialog flow (e.g. determining a point in time). Interpretation results have to be confirmed by the user. In case the answer is not understood, not correct or not confirmed, the question is repeated. Alternatively, for specific questions domain specific actions for clarification are implemented. For example, a choice for specific phone model is shown in Figure~\ref{fig:prototypescreenshot}. For each question, users may ask for details or for an example answer. A skip intent is recognized and causes the dialog to advance to the next question if the current question is optional.

After the questionnaire is concluded, the bot thanks the user and stores the data. In a real-life application, claim management systems would be integrated to automatically trigger subsequent processes.

\section{\uppercase{Evaluation}}
\label{sec:evaluation}

\noindent To evaluate the produced prototype's quality and performance, we conducted a model trial with the goal to report a claim by using the chatbot without having any further instructions available.

Of the 14 participants (who all had some technical background), 35.7\% claimed to regularly use chatbots, 57.1\% to use them occasionally, and only 7.1\% stated that they had never talked to a chatbot before. However, all participants were able to report a claim within a range of about four minutes, resulting in an overall task completion rate of 100\%.

\begin{figure}[hpbt]
\centering
\begin{tikzpicture}[scale=0.85]
  \begin{axis}[
  xscale=0.8,
    ybar,
    enlargelimits=0.15,
    legend style={at={(0.5,-0.2)}, 
      anchor=north,legend columns=-1, ymin=0,ymax=10},
    ylabel={Average rating points},
    symbolic x coords={Ease of Use,Appropriate Formality,Natural Interaction,Response Quality,Personality,Funny \& Interesting, Entertainment
		},
    xtick=data,
    nodes near coords, 
	nodes near coords align={vertical},
    x tick label style={rotate=45,anchor=east},
    ]
    \addplot coordinates {(Ease of Use,8.0) (Appropriate Formality,8.3) 
		(Natural Interaction,7.9) (Response Quality,7.6) (Personality,5.2) (Funny \& Interesting,7.2) (Entertainment,7.7) 
		};
  \end{axis}
\end{tikzpicture}
\caption{Survey results: average user experience ratings (fourteen participants, 0..10 points).}\label{fig:eval}
\end{figure}
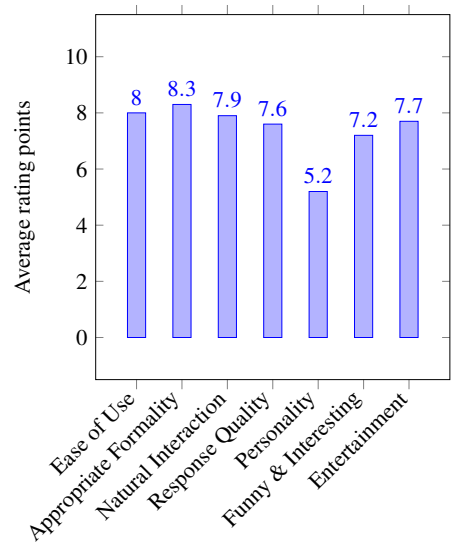

Additionally, the users had to rate the quality of their experiences with the conversational agent by filling out a questionnaire. For each question they could assign points between 0 (did not apply at all) and 10 (did apply to the full extent). The most important quality criteria, whose choice was oriented on the work of~\cite{evaluating2017}, are listed with their average ratings in Figure \ref{fig:eval} and are discussed in detail.

With an average of 8 points for \emph{Ease of Use}, the users had no problems with using the bot to solve the task. In the same way, 8.3 points for \emph{Appropriate Formality} indicate that the participants were comfortable with the formal and informal language the bot talked to them. Only one user stated that he felt worried about permanently being called by his first name after he told it. Fewer points were given for the bot's degree of human-like behavior:
The rating for convincing \emph{Natural Interaction} with 7.9 points may be due to the fact that the conversation was designed in a strongly questionnaire-oriented way, which might have restricted the feeling of having a free user conversation. Also, the satisfaction with given answers to users' domain specific questions was considered quite (but not totally) convincing with 7.6 points. The least convincing experience was that chatbot's \emph{Personality}, which was rated with only 5.2 points on average. This is not surprising, since during this work we put comparatively less efforts in strengthening the agent's personal skills as it does not even introduce itself with a name, but instead mainly acts on a professional level, always concentrating on the fulfillment of its task. With 7.2 points, talking to the chatbot was experienced as quite \emph{Funny \& Interesting}, but still with a lot of room for further improvement. Similarly, the agent's \emph{Entertainment} capabilities, which are at 7.7 points on average at the moment, could be upgraded by extending the conversational contents with additional enjoyable features not related to the questionnaire. For the future we plan to do another larger evaluation on a bigger and more heterogeneous group of participants.

\section{\uppercase{Conclusions and Outlook}}
\label{sec:futurework}

\noindent In this work we have investigated the potential of using conversational agents in insurance companies by general research and implementing a prototype. We determined which classes of conversational agents are useful for insurance companies, which insurance processes can be supported, and what the requirements and motivations for using conversational agents in insurance are. These findings can be used to facilitate the development of conversational agents. We found a need for \emph{Specialized Digital Assistants} in customer facing processes.

Based on these findings we formulated requirements for conversational agents in insurance and selected the smartphone damage claim as an example scenario. We implemented this scenario in a prototype, using machine learning for intent recognition but relying on manual dialog design. Instead of a single dialog state, we implemented a system of multiple conversational states enabling more flexible conversations. We evaluated our prototype with real users and gathered their reactions with a questionnaire. Overall, we found that the prototype is able to handle the example scenario to the user's satisfaction. Possible improvements in the prototype scenario are a better determination of the desired degree of formality as well as defining a consistent persona for the agent (a first step would be to provide a name).

The findings indicate technology is ready to implement conversational agents for insurance customer service scenarios. However, as real-life scenarios are broader than the example scenario, considerably more effort is necessary to design the dialogs. One area not covered by the prototype is human handover in case the conversational agent cannot complete and interaction to the user's satisfaction. 

Data protection and privacy remain open areas for research. Legal and practical questions regarding data collection, storage and processing must be worked on alongside technical requirements, as they tend to be complex and have limited precedent~\citep{smartdatarecht2018}.

In future research, we would like to extend the prototype to different scenarios as well as perform a real-life evaluation with an insurance partner to quantify the benefits of agent use, e.g. call reduction, success rate, and customer satisfaction.

\vfill
\bibliographystyle{apalike}
{\small
\bibliography{./bibliography/bibliography}}

\vfill
\end{document}